\documentclass[preprint,showpacs,preprintnumbers,amsmath,amssymb]{revtex4}

\usepackage{graphicx}
\usepackage{dcolumn}
\usepackage{bm}

\begin{document}

\preprint{APS/123-QED}

\title{Understanding the electroluminescence emitted by single molecules in scanning
tunneling microscopy experiments}

\author{John Buker}
\affiliation{Physics Department, Simon Fraser University, Burnaby, British Columbia, Canada V5A 1S6}
\author{George Kirczenow}%
\affiliation{Physics Department, Simon Fraser University, Burnaby, British Columbia, Canada V5A 1S6}%

\date{\today}

\begin{abstract}

We explore theoretically the electroluminescence of single molecules. We 
adopt a local-electrode framework that is appropriate for scanning 
tunneling microscopy (STM) experiments where electroluminescence 
originates from individual molecules of moderate size on complex 
substrates: Couplings between the STM tip and molecule and between the 
molecule and multiple substrate sites are treated on the same footing, 
as local electrodes contacting the molecule. Electron flow is modelled 
with the Lippmann-Schwinger Green function scattering technique. The 
evolution of the electronic energy levels of the molecule under bias is 
modelled assuming the {\em total} charge of the molecule to be 
invariant, consistent with Coulomb blockade considerations, but that the 
electronic occupations of the molecular HOMO and LUMO levels vary with 
changing bias.  The photon emission rate is calculated using Fermi's 
golden rule. We apply this theoretical approach to the 
STM/Zn-etioporphyrin/Al$_2$O$_3$/NiAl(110) system, and simulate various 
configurations of coupling strength between the molecule and substrate.  
We compare our results to the experimental observations of Qiu, Nazin 
and Ho [Science {\bf 299}, 542 (2003)] for this system and find that our 
model provides a comprehensive explanation of a multitude of previously 
unexplained observations. These include the different types of 
current-voltage characteristics (CVC's) that are observed 
experimentally, the observed association of electroluminescence with 
some CVC's and not others and key properties of the observed photon 
spectra. Theoretical predictions are presented for further 
single-molecule electroluminescence experiments.

\end{abstract}

\pacs{Valid PACS appear here}
\maketitle

\section{\label{sec:level1}Introduction}

In the past 15 years, molecular electronics has become a field of 
intense interest for fundamental research, with potential applications 
in the creation of nanoscopic devices \cite{Ratner06, Tao06, GK}. At the 
same time, great progress has been made in the creation of nanoscale 
photonic devices such as those based on photonic band gap 
materials\cite{Lourtioz}.

The scanning tunneling microscope (STM) is proving immensely useful in 
bridging the gap between these two fields. In STM experiments on clean 
surfaces, light emission has frequently been observed: Systems with an 
STM tip over a metallic\cite{Coombs88, Berndt91} or 
semiconducting\cite{Abraham90, Downes98} surface are known to give off 
light due to the decay of plasmons.  Enhanced photon emission has been 
observed when molecules are placed inside the tip-substrate 
junction\cite{Khaikin90, Berndt93, Flaxer93, Poirier01, Touhari01, 
Sakamoto02}. However, it was unclear until recently whether the stronger 
emission was limited to an amplification of the plasmon-based emission 
seen on metallic surfaces\cite{Barnes98, Hoffmann02}, or whether there 
could in some cases be a different, inherently molecular emission 
mechanism at work.

Recently, it has been definitively demonstrated through STM experiments 
that electric current flow through a molecule may indeed cause the 
molecule to luminesce\cite{Ho03, Dong04} due to molecular orbital 
electronic transitions. This phenomenon, bridging the areas of photonics 
and molecular electronics, is a promising step towards an emerging field 
of single-molecule optoelectronics.

Much insight into the electronic properties of these 
STM/molecule/substrate systems has been obtained by directly studying 
electric current, for example through the comparison of experimental and 
theoretical STM topographs and current-voltage (I-V) curves. In recent 
years, with photonic properties of STM/molecule/substrate systems also 
being studied, a photon detector and spectrometer have been added to the 
standard STM apparatus. Simultaneous photon emission and electric 
current measurements have the potential to greatly enhance our 
understanding of these systems. A theoretical understanding of single 
molecule electroluminescence, however, is still in the earliest stages 
\cite{Buker02, Harbola, Galperin0506} and contact between theory and any 
specific experiment has not been made. The purpose of this article is to 
begin to bridge this divide between theory and single-molecule 
electroluminescence experiments.
                             
The basic idea of molecular electroluminescence as observed in STM 
experiments is as follows: By positioning an STM tip above a single 
molecule on a substrate, and applying a bias voltage between the tip and 
substrate, electron transmission through the molecule may occur, 
mediated by the molecule's electronic orbitals, and the molecule may be 
found to luminesce. In a simplified picture, when a bias voltage is 
applied, the molecule moves out of equilibrium, with a flux of electrons 
passing through it. If two molecular orbitals are located in the energy 
window between the electrochemical potentials of the STM tip and 
substrate, they will both be partially occupied and, if optical 
transitions between them are not forbidden, transitions from the 
higher-energy orbital to the lower-energy orbital will occur, resulting 
in photon emission\cite{Buker02}.

It has been predicted \cite{Buker02} and confirmed experimentally 
\cite{Dong04} that the {\em relative} coupling strengths of the molecule 
to the electron source and drain greatly affect molecular 
electroluminescence. If the coupling strengths are highly asymmetric, 
photon emission is severely quenched.\cite{Buker02} Thus, in 
STM/molecule/substrate experiments where the STM-molecule coupling is 
normally weak, a thin insulating `spacer' layer between the molecule and 
metallic substrate can enhance photon emission by reducing the strength 
of the molecule-substrate coupling and making it comparable with the 
molecule-STM coupling. Conversely, this spacer layer has also been shown 
to strongly suppress plasmon-mediated photon emission, and thus 
facilitate resolving molecular electroluminescence from the background 
plasmon-mediated photon emission that may be present even in the absence 
of a molecule between the STM tip and substrate. For instance, Qiu, 
Nazin and Ho have studied electroluminescence of Zn(II)-etioporphyrin I 
using a 5\AA-thick aluminum oxide insulating layer below the molecule 
(STM/Zn-etioporphyrin/Al$_2$O$_3$/NiAl(110))\cite{Ho03}. In these 
experiments, the STM image, the measured I-V curve and the observation 
of molecular electroluminescence all depend on the precise location of 
the molecule on the Al$_2$O$_3$/NiAl(110) substrate.

In order to theoretically model systems such as these, where there is a 
thin insulating `spacer' layer\cite{plasmonfoot} that has a complex 
atomic structure and a local geometry under the molecule that is not 
measured experimentally but transmits electrons nonuniformly, a local 
electrode approach has proved useful \cite{Buker05}. By considering the 
tip-molecule and molecule-substrate couplings on equal footings, as {\em 
local} electrodes coupled to the molecule, it has been shown that the 
experimentally observed location-dependent STM images of the molecules 
can be explained in terms of different locations of dominant 
molecule-substrate coupling \cite{Buker05}. For the 
STM/Zn-etioporphyrin/Al$_2$O$_3$/NiAl(110) system, there is evidence 
that the out-of-plane ethyl groups of the molecule may be the locations 
of dominant molecule-substrate coupling, and that the {\it strength} of 
the coupling between each ethyl group of the molecule and the substrate 
depends on the location of these groups on the substrate \cite{Buker05}. 
Thus it differs from molecule to molecule adsorbed on the substrate. 
However, to date there has been no theory of electroluminescence from 
this system.

In this article, we extend the above local-electrode theoretical 
framework to the study of electroluminescence and I-V characteristics 
observed in the experimental system of 
STM/Zn-etioporphyrin/Al$_2$O$_3$/NiAl(110).\cite{Ho03} We consider one 
local STM tip probe above the molecule, and four local substrate 
contacts positioned below the four ethyl groups of the molecule. By 
varying the coupling strengths between the molecule and each of the 
electrodes, differing configurations can be simulated. In this model, 
each electrode is represented using a one-dimensional tight-binding 
model, and electron flow is modelled using the Lippmann-Schwinger 
Green-function scattering technique. Fermi's Golden Rule is used to 
calculate photon emission spectra.

At the present time, there is no satisfactory first principles theory of 
the electronic structure of molecules that are weakly coupled to the 
electrodes under applied bias \cite{Chang08}, the situation under 
consideration here: The {\em ab initio} approach to electrical 
conduction based on standard time-independent density functional theory 
breaks down for such systems, yielding unphysical behavior for the 
molecular energy levels and the transmission resonances associated with 
them as the applied bias is varied, and therefore incorrect calculated 
current-voltage characteristics for the molecule \cite{Chang08}. Thus, 
we adopt a different theoretical approach: We use semi-empirical 
extended H\"{u}ckel parameters \cite{Ammeter,Yaehmop} to calculate the 
molecular orbitals and their energies at zero applied bias. The 
dependence of the molecular energy levels on the applied bias is then 
calculated by a self-consistent procedure based on the assumption that 
the net charge on the molecule does not change significantly as the bias 
applied between the STM tip and substrate is varied in the range of bias 
voltages being considered. This assumption is known to be appropriate 
for molecules weakly coupled to the electrodes, for example, in the 
Coulomb blockade regime that is not captured correctly by density 
functional theory. For the present system this methodology is remarkably 
successful, and we are able to attribute the prominent features of the 
experimental data (all of the peaks in the differential conductance vs. 
applied bias, the bias voltages at the onset of electroluminescence and 
the energies of peaks in the observed photon spectra) to the movement of 
the molecular LUMO and HOMO energy levels relative to the 
electrochemical potentials of the source and drain electrodes that 
follows directly from the requirement that the charge on the molecule is 
approximately independent of the applied bias.

The experimental conductance and electroluminescence data for this 
system is multifaceted, depending qualitatively on the location of the 
molecule being probed on the Al$_2$O$_3$ substrate\cite{Ho03} that has a 
complex microscopic structure.\cite{Kresse} In order to account for all 
of the data, we find that it is necessary to include the possibility of 
breaking of the fourfold symmetry of the Hamiltonian of the isolated 
molecule in the model. This is done phenomenologically in two different 
ways: In one of these (Approach A) it is assumed that the symmetry is 
broken by the interaction of the molecule with the complex substrate. In 
the other (Approach B) it is assumed that the symmetry is broken through 
the application of bias between the STM tip and substrate. We find that 
both approaches are generally successful but that Approach A is able to 
better account for one of the features of the experimental data than 
Approach B. At present, since only one experiment of this kind is 
available, it is difficult to judge whether this difference between the 
two approaches confers a substantial advantage to one of them over the 
other. The approaches do, however, offer different predictions for 
experiments that have not yet been carried out.

We find that photon emission is sensitive to the details of the 
molecule-substrate coupling, consistent with experimental data and the 
local-electrode interpretation of the experimental system. We also 
present calculated I-V characteristics for various coupling 
configurations and examine the relationship between the features found 
in the I-V characteristics and the occurrence and nature of the 
luminescence emitted by the molecule. Experimentally, photon emission 
was found to occur when there are two peaks in dI/dV. We find that for 
some coupling configurations, photon emission is predicted and the 
characteristic two-peak curve is obtained. For another configuration, 
only one peak in dI/dV is obtained and photon yield is very low. This is 
also in good qualitative agreement with experiment. Finally, we present 
a case of very weak molecule-substrate coupling, that has not yet been 
achieved experimentally, in which relatively high quantum efficiencies 
are predicted for photon emission.

The organization of this article is as follows: In Section \ref{Model} 
we describe our model and our method of solution. In Section 
\ref{Results} we present our results, compare them with the experimental 
data of Qiu, Nazin and Ho \cite{Ho03} and offer some predictions that 
may be tested in future experiments. In Section \ref{Conclusions} we 
present a concise summary of the aspects of the experimental data that 
our theory has been able to explain and of the physical mechanisms that 
we have identified as being responsible for them. We also comment 
further on the significance of the present work for the fields of 
single-molecule electronics and optoelectronics.

\section{The Model}
\label{Model}

The present model is a generalization of the simpler models presented in 
Refs. \onlinecite{Buker02} and \onlinecite{Buker05}. In the present 
model, the tip and substrate are represented by a tip electrode (probe) 
and substrate electrodes (contacts), each modelled as one-dimensional 
tight-binding chains. Unlike in Refs. \onlinecite{Buker02} and 
\onlinecite{Buker05} where the formalism only allows single substrate 
contacts, in the formalism presented here an arbitrary number of 
substrate contacts are allowed. (We consider cases of 4 substrate 
contacts in the Results section of this article.) The roughly planar 
molecule lies on the substrate and is positioned between the tip and 
substrate electrodes, so that it mediates electron flow between the tip 
and substrate. The electronic model Hamiltonian for this system can be 
divided into three parts, $H=H_{electrodes}+H_{molecule}+W$, where $W$ 
is the interaction Hamiltonian between the electrodes and the molecule. 
Generalizing the Hamiltonian of Ref. \onlinecite{Buker05} to allow 
multiple substrate contacts, the Hamiltonian for the electrodes is given 
by
\begin{align} 
H_{electrodes} &= \sum_{n=-\infty}^{-1}\epsilon|n\rangle\langle 
n|+\beta(|n\rangle\langle n-1|+|n-1\rangle\langle n|)\nonumber\\ 
&+\sum_{i=1}^m\sum_{n=1}^{\infty}\epsilon|n,i\rangle\langle n,i|+\beta 
(|n,i\rangle\langle n+1,i|+|n+1,i\rangle\langle n,i|), 
\label{Helectrodes} 
\end{align} 
where $\epsilon$ are the site energies for the electrodes, $\beta$ is 
the hopping amplitude between nearest-neighbor electrode atoms 
\cite{vcancel}, and $|n\rangle$ and $|n,i\rangle$ represent orbitals at 
site $n$ of the tip probe and site $n$ of the $i$th substrate contact, 
respectively. We take the electrochemical potentials of the tip and 
substrate electrodes to be $\mu_T=E_F+eV_{bias}/2$ and 
$\mu_S=E_F-eV_{bias}/2$, where $V_{bias}$ is the bias voltage applied 
between them and $E_F$ is their common Fermi level at zero bias. The 
Hamiltonian of the molecule may be expressed as
\begin{equation}
H_{molecule}= 
\sum_{j}\epsilon_j|\phi_j\rangle\langle\phi_j|, \label{Hmol} 
\end{equation}
where $\epsilon_j$ is the energy of the $j$th molecular 
orbital ($|\phi_j\rangle$). Unlike in Ref. 
\onlinecite{Buker05}, molecular orbital energies are allowed to shift in 
response to an applied bias voltage. Our treatment of the effect of bias 
voltage on orbital energies is described in Sec. IID. The interaction 
Hamiltonian between the electrodes and molecule is given by 
\begin{equation} 
W = \sum_{j}(W_{-1,j}|-1\rangle\langle\phi_j| + 
W_{j,-1}|\phi_j\rangle\langle -1| +\sum_{i=1}^m 
[W_{j,(1,i)}|\phi_j\rangle\langle 1,i| 
+W_{(1,i),j}|1,i\rangle\langle\phi_j|]),
\label{Hint}
\end{equation} 
where $W_{-1,j}$, $W_{j,-1}$, $W_{j,(1,i)}$, and $W_{(1,i),j}$ are the 
hopping amplitude matrix elements between the electrodes and the various 
molecular orbitals $|\phi_j\rangle$.

Electrons propagate in the form of Bloch waves through each electrode 
toward the molecule, and may undergo transmission or reflection when 
they encounter the molecule, contributing to the occupation of molecular 
orbitals in the process. Wave functions of electrons incoming from the 
tip probe are of the form
\begin{equation}
|\psi\rangle=\sum_{n=-\infty}^{-1}(e^{iknd} +
re^{-iknd})|n\rangle+\sum_{i=1}^m\sum_{n=1}^{\infty}t_i e^{ik
nd}|n,i\rangle+\sum_{j}c_{j}|\phi_{j}\rangle
\label{psi}
\end{equation}
where $d$ is the lattice spacing, $t_i$ are the transmission 
coefficients into the different substrate contacts, and $r$ is the 
reflection coefficient.

\subsection{Solving the system}
In order to calculate molecular-based photon emission and I-V 
characteristics, it is necessary to evaluate the molecular orbital 
coefficients and transmission amplitudes for incoming electrons. This 
may be done by solving a Lippmann-Schwinger equation for this system, in 
a similar fashion to Ref. \onlinecite{Buker05} but generalized to 
multiple substrate contacts:
\begin{equation}
|\psi\rangle=|\phi_{0}\rangle+G_{0}(E)W|\psi\rangle,
\label{lippmann}
\end{equation}
where $G_0(E)=(E-(H_{electrodes}+H_{molecule})+i\delta)^{-1}$ is the 
Green function for the decoupled system (without $W$), and 
$|\phi_0\rangle$ is the eigenstate of an electron in the decoupled tip 
probe (or, more generally, the incoming electrode). $G_0(E)$ may be 
separated into the decoupled components: the tip and substrate 
electrodes, and the molecule. For each electrode,
\begin{equation}
G_0^{electrode} = \sum_k\frac{|\phi_0(k)\rangle\langle\phi_0(k)|}
{E-(\epsilon+2\beta cos(kd))}
\label{gprobe}
\end{equation}
where $d$ is the lattice spacing and $\epsilon+2\beta cos(kd)$ is the 
energy of an electrode electron with wave vector $k$. $G_0^{electrode}$ 
may also be expressed in an atomic orbital basis:
\begin{equation}
G_0^{electrode}=\sum_{n=1}^{\infty}\sum_{m=1}^{\infty}
(G_0^{electrode})_{n,m}|n\rangle\langle m|,
\label{GRatom}
\end{equation}
whose matrix elements $(G_0^{electrode})_{n,m}$ are known 
analytically.\cite{Emberly98} For the molecule,
\begin{equation}
G_0^M=\sum_j\frac{|\phi_j\rangle\langle\phi_j|}{E-\epsilon_j}
=\sum_j(G_0^M)_j|\phi_j\rangle\langle\phi_j|.
\label{gmolecule}
\end{equation}
For an electron incoming from the tip probe, this yields the following 
set of linear equations for the coefficients of 
$|\psi\rangle$:\cite{Emberly98}
\begin{align}
\psi_{-1}&=(\phi_0)_{-1}+(G_0^{electrode})_{-1,-1}\sum_j W_{-1,j}c_j\\
\psi_{1,i}&=(G_0^{electrode})_{1,1}\sum_j W_{(1,i),j} c_j\\
c_j&=(G_0^M)_j(W_{j,-1}\psi_{-1}+W_{j,(1,i)}\psi_{1,i})
\label{gmatrix}
\end{align}
where $\psi_{-1}=\langle-1|\psi\rangle$, 
$\psi_{1,i}=\langle 1,i|\psi\rangle$, and 
$(\phi_0)_{-1}=\langle-1|\phi_0\rangle$. The transmission probability 
for an electron incoming from the tip probe is given by 
$T=\sum_{i=1}^m\frac{v^\prime}{v}|t_i|^2.$\cite{vcancel} Tip probe 
electrons between $\mu_T$ and $\mu_S$ in energy contribute to the 
electric current through the molecule. Using the Landauer 
theory,\cite{Landauer} an expression for the current is obtained:
\begin{equation}
I=\frac{2e}{h}\int_{\mu_S}^{\mu_T}T(E,V_{bias})dE.
\label{Landauercurrent}
\end{equation}
The dependence of $T$ on $V_{bias}$ is due to shifting molecular orbital 
energies (described in Sec. IID). 

\subsection{Photon emission}
Photon emission from the molecule can be understood in terms of allowed 
electronic transitions from a molecular orbital to one with a lower 
energy. To calculate emission spectra, as in Ref. \onlinecite{Buker02} 
we use the expression for the spontaneous emission rate of a system 
emitting photons into empty space, using Fermi's Golden 
Rule.\cite{Ballentine} The emission rate is given by
\begin{equation}
\frac{4e^2\omega^3}{3\hbar c^3}|\langle\psi_f|{\bf x}|\psi_i\rangle|^2,
\label{FermiRate}
\end{equation}
where $\psi_i$ and $\psi_f$ represent initial and final states, and 
$\hbar\omega$ is their difference in energy. Unlike in Ref. 
\onlinecite{Buker02}, where photon emission is calculated for the 
idealized case of a two-orbital molecule, here we calculate photon 
emission for a system involving a real molecule with multiple molecular 
orbitals. In order to do this, we consider emission only from the 
molecule itself. The rate is therefore approximated by
\begin{equation}
R(k_i, \omega)=\frac{4e^2\omega^3}{3\hbar c^3}
|\sum_{j,j^\prime}|c_{j^\prime,f}|^2|c_{j,i}|^2
|\langle\phi_{j^\prime}|{\bf x}|\phi_j\rangle|^2,
\label{RateApx} 
\end{equation}
where $i$ and $f$ label initial and final states. The relevant 
transition dipole moments $\langle\phi_{j^\prime}|{\bf x}|\phi_j\rangle$ 
are calculated by performing an extended H\"{u}ckel dipole analysis of 
the molecular orbitals.\cite{iconedit} To calculate the emission rate as 
a function of photon energy, we generalize the procedure presented in 
Ref. \onlinecite{Buker02}. We must consider all electron states of the 
system, incoming from both the tip probe and each of the substrate 
contacts. Each electron state consists of an incoming wave, transmitted 
wave, reflected wave, and an amplitude on the molecule. See 
Fig.~\ref{fig1} for a schematic illustration. We assume here (and 
throughout the article) the positive bias case, with $\mu_T > \mu_S$. 
Since we assume the temperature to be 0 K, all states with incoming 
waves from a given electrode are occupied up to the electrochemical 
potential of that electrode. For a transition to occur, $\psi_f$ must be 
an unoccupied state, and it must be lower in energy than $\psi_i$. 
Therefore, we consider transitions from occupied initial states (below 
$\mu_T$) that are incoming from the tip probe, to unoccupied final 
states within the electrochemical potential window (above $\mu_S$) that 
are incoming from one of the substrate contacts. After normalizing the 
wave functions and converting the sum over $k$ states (and spin) into an 
integral over energy, an expression for the photon emission spectrum 
(for a given bias voltage) is obtained:
\begin{equation} 
f(\omega)=\frac{1}{2\pi}\sum_{contacts}\int_{\mu_S+\hbar 
\omega}^{\mu_T}\frac{R(k_i,\omega)}{-\beta sin(k_i d)}dE_i, 
\label{Spectrum} 
\end{equation}
where $E_i$ and $k_i$ are the initial energy and wave vector of an 
electron incoming from the tip probe, and $\omega$ is the frequency of 
the photon emitted.

\subsection{Electronic structure of Zn-etioporphyrin at zero bias} The 
electronic structure of the Zn-etioporphyrin molecule was computed using 
the extended H\"{u}ckel model.\cite{Yaehmop} Within this model, the 
energy of the highest occupied molecular orbital (HOMO) is -11.5 eV, and 
the energy of the LUMO is -10.0 eV. For a charge neutral molecule at 
equilibrium, weakly coupled to the electrodes, the Fermi level of the 
electrodes at zero bias is expected to be located between the molecular 
HOMO and LUMO levels. However, the precise location of the Fermi level 
is a difficult problem in molecular electronics, with differing 
theoretical approaches yielding differing results. In STM experiments on 
Zn-etioporphyrin, the appearance of a low-bias dI/dV peak for some 
positions of the molecule above the substrate implies a Fermi level that 
is close to either the HOMO or LUMO. Since these are STM experiments, it 
is likely that the low-bias peak corresponds to an orbital entering the 
Fermi energy window by crossing $\mu_T$ (the electrochemical potential 
of the STM tip) rather than $\mu_S$, due to the weaker coupling of the 
orbital to the tip than to the substrate. Since these experiments were 
performed at positive bias (electron flow from tip to substrate), it is 
therefore likely that the Fermi level is close to the LUMO and not the 
HOMO.

The precise location of the Fermi level of the electrodes below the LUMO 
energy is likely to depend on the local geometry of the 
Zn-etioporphyrin/Al$_2$O$_3$/NiAl(110) interface: A work function 
study\cite{Song05} of Al$_2$O$_3$ on NiAl(110) has found that the 
formation of an ultrathin Al$_2$O$_3$ layer on NiAl(110) decreases the 
work function of the substrate by about 0.8 eV, with a strong dependence 
on the oxide layer structure and thickness. It is therefore reasonable 
to assume that, for the experiments by Qiu, et al.,\cite{Ho03} different 
locations on the Al$_2$O$_3$/NiAl(110) substrate have different local 
work functions, with differences on the order of a few tenths of an eV. 
Due to these differences, variations in the common zero-bias Fermi 
energy of the tip and substrate (relative to the vacuum and also to the 
energies of the molecular orbitals) are likely to occur. Our 
calculations show that in most cases the overall qualitative picture is 
not sensitive to the precise location of the Fermi level below the LUMO 
energy. Therefore, for a qualitatively reasonable analysis of this 
system, we choose a zero-bias Fermi level of -10.1 eV. We justify this 
reasoning more explicitly in Sec. IIIB(3), where we compare our results 
for this Fermi level with results obtained assuming a zero-bias Fermi 
level of -10.3 eV.

\subsection{Molecular orbital energy-level dependences}
In order to realistically model photon emission and electric current as 
a function of bias voltage, it is necessary to consider the effects of 
bias voltage on molecular orbital energies. When a bias voltage is 
applied, an electric field is created between the tip and substrate, 
which may result in some charging of the molecule. If this occurs, the 
charging causes an electrostatic shift of the molecular energy levels 
that in turn severely limits the actual charging that takes 
place.\cite{Emberly01} Generalizing the minimal charging approximation 
presented in Ref. \onlinecite{Buker02}, we phenomenologically 
approximate the shift of the molecular levels in response to the applied 
bias by adjusting $\epsilon_j$ for each molecular orbital so as to 
maintain the net charge that the molecule has at zero bias. The net 
electronic charge is calculated by summing over all occupied electron 
states (including spin) incoming from each electrode. This sum is 
converted into an integral, and an expression for the charge is 
obtained:
\begin{equation} 
Q=\frac{1}{2\pi}(\int^{\mu_T} \sum_j 
\frac{|c_j(E,V_{bias})|^2}{-\beta sin(kd)}dE +\sum_{contacts} 
\int^{\mu_S} \sum_j \frac{|c_j(E,V_{bias})|^2}{-\beta sin(kd)}dE), 
\label{molcharge} 
\end{equation} 
where the molecular orbital energies $\epsilon_j$ (and therefore $c_j$) 
change with $V_{bias}$ in such a way that $Q = constant$. 

\subsubsection{Approach A} One approach to treating the bias dependence 
of molecular orbital energies $\epsilon_j$ (which we will call Approach 
A) is to assume an equal bias dependence for the shifts in energy of 
each molecular orbital. This simple approach to charging yields 
physically reasonable behaviour of the molecular orbital energy levels 
with bias. However, by itself, it is insufficient to explain many of the 
experimentally observed STM I-V characteristics and photon emission 
results for the molecule Zn-etioporphyrin on Al$_2$O$_3$/NiAl(110). This 
may be because Zn-etioporphyrin has a two-fold degenerate lowest 
unoccupied molecular orbital (LUMO) that is likely to lose its 
degeneracy when the molecule is placed on a region of the complex 
surface where the molecule-substrate interaction is not 
fourfold-symmetric. After including such a substrate-dependent splitting 
in the zero-bias electronic structure of the LUMO, this approach yields 
interesting results that are consistent with the experimental data.

\subsubsection{Approach B} Another approach (Approach B) is to consider 
the bias dependence of the different orbital energies in a slightly more 
complex way. Since Zn-etioporphyrin is a planar molecule and all of the 
relevant orbitals except for the LUMO have fourfold symmetry, the 
electric field from the STM tip affects each of the fourfold symmetric 
orbitals similarly, and we adjust their energies by equal amounts 
$\alpha$. The LUMO, however, consists of two degenerate orbitals with 
twofold symmetry. Depending on the position of the STM tip above the 
molecule, as bias voltage is applied this may result in a stronger 
electric field effect on the energy of one of the LUMO orbitals, and a 
weaker effect on the other orbital. Therefore, for cases where the tip 
probe is positioned above a region of the molecule with a high amplitude 
for one LUMO orbital and a low amplitude for the other, instead of 
adjusting the LUMO energies by equal amounts $\alpha$ we adjust the LUMO 
energies by amounts of $\gamma_1$ ($>\alpha$) and $\gamma_2$ ($<\alpha$) 
respectively. Within the present model, the quantities $\alpha$, 
$\gamma_1$ and $\gamma_2$ all change with $V_{bias}$ such that the total 
molecular charge $Q$ remains constant. These quantities depend on the 
electrostatic geometry of the system. Therefore, for all values of 
$V_{bias}$, the ratios $\alpha:\gamma_1:\gamma_2$ are kept the same, 
consistent with the linearity of electrostatics. With this 
phenomenological approach to charging, unlike in Approach A, we do not 
assume there to be any zero-bias splitting of the LUMO degeneracy.

In the remainder of this paper we will present photon emission results 
and current-voltage (I-V) characteristics for Zn-etioporphyrin, 
calculated based on the above model for both Approach A and Approach B. 
We will show how photon emission is sensitive to details of the 
molecule-substrate coupling, and explore the relationship between photon 
emission and I-V curve features. In addition, we will demonstrate how 
our model can account for many previously unexplained features of the 
experimental data for this system.

\section{Results}
\label{Results}

We present results for Zn(II)-etioporphyrin I, coupled to a tip probe 
and 4 substrate contacts that we represent for simplicity by Cu $s$ 
orbitals. The geometrical structure of the molecule has been calculated 
using density-functional theory.\cite{DFT} The molecule is mainly planar 
and oriented approximately parallel to the substrate, but contains four 
out-of-plane ethyl groups.

\subsection{Strong fourfold-symmetric molecule-substrate coupling} We 
first consider a case where there is strong electronic 
molecule-substrate coupling relative to the coupling between the 
molecule and the STM tip, and where the molecule-substrate interaction 
is fourfold-symmetric. By `strong coupling' we mean that the Hamiltonian 
matrix elements $W_{electrode,j}$ between the relevant molecular 
orbitals and substrate contacts are about an order of magnitude greater 
than between the molecular orbitals and tip probe. It has been 
previously shown that the out-of-plane ethyl groups of the molecule are 
likely locations of dominant molecule-substrate coupling.\cite{Buker05} 
Therefore, four local substrate contacts (S1-S4) are positioned below 
the ethyl groups of the molecule, as shown in 
Fig.~\ref{fig2}.\cite{strongfoot} For Approach A (described in Sec. 
IID(1)), in this case we assume there is no splitting of the LUMO 
degeneracy, consistent with the fourfold symmetry of the 
molecule-substrate coupling. The tip probe is positioned (see 
Fig.~\ref{fig2}) above the molecule in a lateral region that has been 
shown to be part of the observed high-transmission lobe pattern for the 
STM tip above Zn-etioporphyrin.\cite{Ho03,Buker05} For this position of 
the tip probe (and any position corresponding to an experimentally 
observed high-transmission lobe) the tip probe has a stronger {\it 
electrostatic} coupling to one of the degenerate twofold symmetric LUMOs 
than to the other, and an intermediate coupling to all other relevant 
orbitals. [The difference between electrostatic and electronic coupling 
should be noted: {\it Electrostatic coupling} refers to the change in 
the electrostatic potential that an electron in a molecular orbital 
feels due to the applied bias voltage, whereas {\it electronic coupling} 
refers to the Hamiltonian matrix element $W_{electrode,j}$ between an 
electrode and a molecular orbital. In the rest of this article, these 
terms will be frequently used.] Therefore, for Approach B (discussed in 
Sec. IID(2)), in order to model the shift of molecular orbital energies 
due to electrostatic effects in a phenomenological, qualitatively 
reasonable way, we assume the ratio $\alpha:\gamma_1:\gamma_2$ 
(discussed in Sec. IID) to be 3:4:2. Here, $\gamma_1$ corresponds to the 
LUMO orbital that has stronger electrostatic coupling, and $\gamma_2$ to 
the orbital that has weaker electrostatic coupling to the tip. Results 
presented throughout this article are not sensitive to the precise 
values chosen for this ratio.\cite{RatioNote}

\subsubsection{Approach A} 
For this strong substrate coupling case, with Approach A, photon 
emission is computed to be very weak. (In Sec. IIIB,C, cases will be 
presented where the photon yield is more than an order of magnitude 
greater.) This weak emission result is consistent with the quenching of 
emission due to asymmetric coupling of the molecule to the tip and 
substrate observed experimentally\cite{Barnes98,Hoffmann02} and 
predicted for the general case of current-carrying molecular 
wires\cite{Buker02}. The quenching of photon emission due to asymmetry 
of the electronic coupling can be understood physically as follows: 
Looking at Fig.\ref{fig1}, in a highly asymmetric system where the 
tip-molecule coupling is much weaker than the molecule-substrate 
coupling, electrons incoming from the tip have relatively low amplitudes 
for entering the molecule, and high amplitudes for exiting into the 
substrate. There is therefore a low amplitude $c_{j,i}$ for an electron 
in its initial state to be on a molecular orbital (even if the orbital 
is inside the Fermi energy window and close in energy to the energy of 
the electron), resulting in a low photon emission rate (see 
Eq.\ref{RateApx}).

A further possible consideration is the molecular orbital amplitude 
$c_{j,f}$ of an electron in its final state. If no allowed molecular 
orbitals are available to receive transitions (ie. inside the Fermi 
energy window of the system), $c_{j,f}$ will be small for all possible 
final states and emission will be further quenched. As we will now show, 
for the strong fourfold-symmetric coupling situation we consider here, 
this in fact is the case.

This further quenching, as well as the calculated current-voltage (I-V) 
curve for this case shown in Fig.~\ref{fig3}a, can be understood by 
studying how the molecular orbital energies shift with bias voltage (see 
Fig.~\ref{fig3}c): The LUMO (assumed to be degenerate in this case), 
becomes partially (slightly) occupied at low bias as the tip 
electrochemical potential ($\mu_T$) approaches its energy. This causes 
an electrostatic shift of the molecular orbital energies (discussed in 
Sec. IID). The LUMO then shifts upwards in energy, following $\mu_T$, so 
that the net charge on the molecule is maintained.

The result is the approximately linear I-V curve at low bias in 
Fig.~\ref{fig3}a, with electron flow being mediated by the tails of the 
HOMO and the LUMO. There is a slight low-bias dI/dV feature due to 
$\mu_T$ approaching the LUMO energy. At about 1.3 V, the slope of the 
I-V curve begins to increase, resulting in a peak in dI/dV. The reason 
for this is as follows: The HOMO begins to become partially (slightly) 
unoccupied, even though it is still below the substrate electrochemical 
potential ($\mu_S$). This is because the molecule-substrate contact 
couplings are strong compared to the molecule-tip coupling, so the 
substrate has a much stronger effect on the orbital occupations than the 
tip, and the high-energy tail of the HOMO begins to depopulate. HOMO 
electrons inside the Fermi energy window contribute to current flow into 
the substrate, increasing the slope of the I-V curve. The orbital 
energies are affected slightly, with the LUMO shifting slightly lower 
relative to $\mu_T$ (but not visibly in Fig.~\ref{fig3}c), such that the 
net charge on the molecule is maintained. The slight downward shift of 
the LUMO energy further increases the slope of the I-V curve. Here, 
electric current is very sensitive to such a shift, due to the LUMO's 
energy being very close to $\mu_T$. At about 1.4 V, the LUMO fully 
enters the Fermi energy window, in the process becoming only slightly 
occupied due to the much weaker coupling of the molecule to the tip 
(electron source electrode) than to the substrate (drain). At this point 
both the HOMO and LUMO orbital energies shift downwards, in such a way 
that the charge on the molecule remains constant (ie. the HOMO energy 
follows $\mu_S$. The HOMO energy remains below $\mu_S$, resulting in 
quenched photon emission. The I-V curve flattens, since no orbitals are 
entering or approaching the energy window between tip and substrate 
Fermi energies.

We now compare this result with experimental results obtained by Qiu et 
al.\cite{Ho03} for the STM/Zn-etioporphyrin/Al$_2$O$_3$/NiAl(110) 
system. In these experiments, depending on the location of the molecule 
on the substrate, the molecule either luminesced or did not, with 
different dI/dV curves obtained for luminescent and non-luminescent 
cases. See Fig.~\ref{fig4} for the reproduced experimental curves. Here, 
curve A and curve B are representative of molecules that were found to 
luminesce. Molecules with current-voltage curves C-F did not exhibit 
observable luminescence. Experimentally, molecules that did not 
luminesce were found to have only one dI/dV peak, usually at around 1.4 
V. This is in good qualitative agreement with the model result presented 
here, using Approach A, which shows only one significant dI/dV peak that 
occurs at 1.4 V in Fig.~\ref{fig3}a, and very weak photon emission, that 
is likely not experimentally detectable.

\subsubsection{Approach B}
For the case of strong fourfold-symmetric molecule-substrate coupling, 
Approach B (discussed in Sec.IID) yields I-V results shown in 
Fig.~\ref{fig3}b that are qualitatively similar to those in 
Fig.~\ref{fig3}a that were obtained using Approach A. Photon emission is 
also computed to be very weak, for the same reasons as with Approach A.

With Approach B, the LUMO with the weaker electrostatic coupling to the 
tip (which we will refer to as LUMO2) enters the Fermi energy window at 
low bias (see Fig.~\ref{fig3}d), but contributes very little to the 
electric current (see Fig.~\ref{fig3}b), due to the very weak LUMO2-tip 
probe electronic coupling. The LUMO2 remains almost completely 
unoccupied because of the asymmetry of the LUMO2-tip and LUMO2-substrate 
couplings. As $\mu_T$ approaches the energy of the {\it more strongly} 
electrostatically and electronically coupled LUMO (LUMO1), however, the 
LUMO1 becomes partially (slightly) occupied and shifts in energy, 
following $\mu_T$, so that the net charge on the molecule is maintained. 
The result is again an approximately linear I-V curve, with electron 
flow being mediated by the tails of the HOMO and the LUMO1.

At about 1.5 V, the HOMO begins to become partially (slightly) 
unoccupied, similarly to Approach A, increasing the slope of the I-V 
curve. The LUMO1 shifts slightly lower relative to $\mu_T$, such that 
the net charge on the molecule is maintained. This further increases the 
slope of the I-V curve. At 1.6 V, the LUMO1 fully enters the Fermi 
energy window, in the process becoming only slightly occupied due to the 
asymmetry of the coupling. As with Approach A, the orbital energies then 
shift downwards, in such a way that the charge on the molecule remains 
constant. The HOMO energy remains below $\mu_S$, resulting in quenched 
photon emission, and the I-V curve flattens.

For this case of strong molecule-substrate coupling using Approach B, 
there is found to be only one significant dI/dV peak (at 1.6 V) and very 
weak photon emission. As for Approach A, this compares well with the 
experimental non-luminescent cases (see Fig.~\ref{fig4}C-F), where one 
dI/dV peak is observed (at about 1.4 V).

\subsection{Localized strong coupling} 

Next, we consider the case where there is strong electronic coupling 
between the molecule and only {\em one} of the four substrate contacts. 
It has been suggested\cite{Buker05} that this type of electrode 
configuration is a likely possibility for the common experimental case 
of Fig.2B in the article by Qiu et al.\cite{Ho03} Significant molecular 
electroluminesce was observed for this experimental case.

The electrode configuration that we consider is similar to Sec.IIIA (see 
Fig.~\ref{fig2}); however, in this case the substrate contacts $S_1$, 
$S_2$ and $S_3$ are moderately coupled to the molecule (coupling less 
than an order of magnitude greater than the coupling to the tip probe), 
and $S_4$ is strongly coupled.\cite{strongmodfoot} The tip probe is 
positioned in the same lateral region as for Sec.IIIA, and again with 
greater electrostatic coupling to one LUMO (LUMO1) relative to the other 
LUMO (LUMO2). It should also be noted that, due to the twofold symmetry 
of the LUMO, the strongly coupled substrate contact is electronically 
strongly coupled to only one of the LUMOs (LUMO2, in this case) and not 
the other (LUMO1).

\subsubsection{Approach A}
With Approach A, since the molecule-substrate interaction is in this 
case not fourfold-symmetric, there is a splitting in the zero-bias 
degeneracy of the LUMO.\cite{splitfoot}

For this case, significant photon emission is computed to occur. 
Fig.~\ref{fig5}a shows the calculated emission spectrum at high bias 
($V_{bias}=1.94$ V). The spectrum corresponds to HOMO-LUMO1 (1.94 eV 
peak) and HOMO-LUMO2 (1.44 eV peak) transitions. The calculated I-V 
curve for this case, shown in Fig.~\ref{fig5}c, has a low-bias dI/dV 
peak and a high-bias dI/dV peak.

To understand the calculated photon emission spectra and I-V curves for 
this case, it is necessary to pay close attention to the details of the 
coupling of the various molecular orbitals to the electrodes. Looking at 
Fig.\ref{fig5}e, at low bias the LUMO2 enters the Fermi energy window, 
remaining almost completely unoccupied due to the strongly asymmetric 
coupling of the LUMO2 to the tip and substrate. In this case, however, 
the LUMO2 contribution to the electric current is not negligible. 
Current flow mediated by the LUMO2 is not drowned out by current flow 
mediated through the tails of the LUMO1 or the HOMO, since in this case 
the electronic coupling of the substrate is strongest to the LUMO2. This 
creates the low-bias dI/dV peak seen in Fig.~\ref{fig5}c.

The LUMO2 energy follows $\mu_T$ up to 0.2 V (see Fig.~\ref{fig5}e). In 
this case, the substrate contacts have a large influence on the 
occupation of the LUMO2 even though the LUMO2 is well above $\mu_S$, 
because the coupling between the substrate and LUMO2 is much stronger 
than between the tip and LUMO2. Thus, from 0.2 V to 0.6 V the LUMO2 
tracks $\mu_S$ and the I-V curve (Fig.~\ref{fig5}c) is flat. At 0.6 V, 
the LUMO1 approaches $\mu_T$ and begins to populate. In response, the 
energies of the orbitals rise such that no charging takes place. The tip 
probe has a large influence on the occupation of the LUMO1, because the 
coupling between the tip/substrate and LUMO1 is not highly asymmetric. 
From 0.6 V to 1.9 V, The LUMO1 and the tail of the HOMO are the dominant 
sources of rising current.

The HOMO reaches $\mu_S$ at $V_{bias}=1.9$ V, causing an electrostatic 
shift in energy of the orbitals downwards, so that the LUMO1 enters the 
Fermi energy window and populates significantly. The HOMO reaches 
$\mu_S$ and depopulates by an equal amount. There is a resulting sharp 
increase in current, as both the HOMO and LUMO1 mediate electron 
transmission from tip to substrate.

Close inspection of Fig.~\ref{fig5}e and Fig.~\ref{fig3}c shows that, in 
this case, the HOMO energy comes up to $\mu_S$ (within the resolution of 
the figure) whereas for the case of Sec. IIIA (Approach A), the HOMO 
energy only {\it approaches} $\mu_S$. Hence, the depopulation of the 
HOMO is much greater in this case than in Sec.IIIA. Greater depopulation 
of the HOMO occurs, along with greater population of the LUMO1 as it 
enters the Fermi energy window, due to the lack of strong coupling 
asymmetry between the LUMO1 and the tip/substrate electrodes in this 
case. In this way, the zero-bias charge is maintained. Therefore, above 
1.9 V, HOMO-mediated electronic states are available to receive 
transitions from LUMO1 and LUMO2-mediated states, resulting in photon 
emission. Since there is a stronger coupling asymmetry between the LUMO2 
and the tip/substrate electrodes than between the LUMO1 and those 
electrodes, LUMO2-HOMO photon emission is weaker than LUMO1-HOMO 
emission (see Fig.~\ref{fig5}a), as explained in Sec.IIIA.

The onset of photon emission in this case occurs as the HOMO becomes 
partially unoccupied, at about 1.9 V. Fig.~\ref{fig6}a shows the onset 
of photon emission, at the spectrum peak corresponding to HOMO-LUMO2 
transitions. Notice that the peak photon energy ($\approx$ 1.43-1.44 eV) 
is significantly less than the Fermi gap energy ($\approx$ 1.93-1.94 
eV). This is because the LUMO2 is deep inside the Fermi energy window at 
the onset voltage (see Fig.~\ref{fig5}c). The calculated photon emission 
peak due to HOMO-LUMO1 transitions has the same onset voltage as the 
HOMO-LUMO2 emission peak. For this transition, however, photon energy is 
peaked close to the Fermi gap energy (1.9 eV) because the LUMO1 and HOMO 
have energies close to $\mu_T$ and $\mu_S$ respectively at the onset 
voltage.

Comparing results for this luminescent case to experiment, the 
similarities are striking. Experimentally, molecules that luminesced 
commonly had a small dI/dV peak at 0.2 V and a larger peak at around 2.0 
V (see Fig.~\ref{fig4}A,B). This is in excellent qualitative agreement 
with Fig.\ref{fig5}c, where we see a small dI/dV peak at 0.2 V and a 
larger peak at about 1.9 V. Furthermore, experimental results\cite{Ho03} 
(reproduced here in Fig.~\ref{fig7}) show the onset of photon emission 
occurring most commonly at about 2.2 V, but with a photon energy peak in 
the spectrum about 0.5 eV below the corresponding Fermi gap energy of 
2.2 eV. This is in good agreement with Fig.~\ref{fig6}, where at onset 
we find an emission peak (corresponding to the HOMO-LUMO2 transition) 
significantly below the Fermi gap energy. Also, comparing 
Fig.~\ref{fig6} with emission onset spectra for the most common 
experimental case (Fig.~\ref{fig7}) we see very similar behaviour of the 
emission spectra tails: The high-energy tail has a sharp cutoff, while 
the low-energy tail does not. As bias voltage increases, the high-energy 
cutoff shifts upwards in energy by a similar amount. In our model, we 
also see this behaviour, because the Fermi energy of the substrate 
provides a sharp energy cutoff below which there are no available final 
states for a transition. This cutoff reduces the extent of the 
high-energy tails. There is no such cutoff reducing the extent of the 
low-energy tails.

Notice also that, experimentally, there is a shift in the position of 
the high-bias dI/dV peak, depending on whether photon emission is 
observed: In Fig.~\ref{fig4} a peak is observed at 1.4 V for 
non-luminescent cases (C-F), and around 2.0 V for luminescent cases 
(A,B). We see the same sort of bias peak shift theoretically with 
Approach A: 1.4 V for Sec.IIIA (weak emission case) and 1.9 V for 
Sec.IIIB (strong emission case). In this way, Fig.~\ref{fig3}a is 
similar to Fig.~\ref{fig4}C-F, while Fig.~\ref{fig5}c is similar to 
Fig.~\ref{fig4}A,B.

Our model further predicts a stronger HOMO-LUMO1 emission peak (the 1.94 
eV peak in Fig.~\ref{fig5}a) with the same onset voltage as the 
experimentally observed HOMO-LUMO2 emission peak, but with a higher peak 
photon energy, close to the Fermi gap energy $=eV_{bias}$. The 
experimental photon spectra in Ref. \onlinecite{Ho03} do not extend to 
the photon energy energy range in which this emission peak is predicted 
to occur (2.2 eV photon energy for the experimental onset voltage of 2.2 
V). An experimental study testing this prediction would be very 
desirable.

\subsubsection{Approach B}
With Approach B, as with Approach A, significant photon emission is 
computed to occur in this case. Fig.~\ref{fig5}b shows the emission 
spectrum at high bias ($V_{bias}=1.94$ V). The spectrum corresponds to 
HOMO-LUMO1 (1.57 eV peak) and HOMO-LUMO2 (1.30 eV peak) transitions. The 
I-V curve for this case, shown in Fig.~\ref{fig5}d, has a low-bias dI/dV 
peak and a high-bias dI/dV peak.

Looking at Fig.\ref{fig5}f, at low bias the LUMO2 enters the Fermi 
energy window. It remains almost completely unoccupied due to the 
strongly asymmetric coupling of the LUMO2 to the tip and substrate, but 
as with Approach A it still contributes to the electric current. This 
results in the low-bias dI/dV peak seen in Fig.~\ref{fig5}d. At 0.2 V 
the energy of the LUMO1 reaches $\mu_T$. This causes an electrostatic 
shift in the energy levels upwards, as shown in Fig.~\ref{fig5}f. From 
0.2 V to 1.6 V, the LUMO1 and the tail of the HOMO are the dominant 
sources of rising current. The HOMO reaches $\mu_S$ at 1.6 V, causing an 
electrostatic shift in energy of the orbitals downwards, so that the 
LUMO1 enters the Fermi energy window and populates significantly. 
Similarly to Approach A, the HOMO reaches $\mu_S$ and depopulates by an 
equal amount, resulting in a sharp increase in current.

For the same reasons as were explained for Approach A, for Approach B at 
1.6 V HOMO-mediated electronic states are available to receive 
transitions from LUMO1 and LUMO2-mediated states, resulting in photon 
emission. As with Approach A, LUMO2-HOMO photon emission is weaker than 
LUMO1-HOMO emission (see Fig.~\ref{fig5}b). Fig.~\ref{fig6}b shows the 
onset of photon emission, around $V_{bias}=1.6$ V, at the spectrum peak 
corresponding to HOMO-LUMO2 transitions. As with Approach A, the photon 
peak energy is significantly less than the Fermi gap energy.

Qualitatively, I-V and photon emission results for Approach B are 
similar to results for Approach A, and compare similarly well to 
experiment. There is one exception: With Approach B, there is no shift 
in the position of the high-bias dI/dV peak depending on whether or not 
photon emission is observed: A peak is predicted at 1.6 V for both 
luminescent and non-luminescent cases, due to the very similar molecular 
orbital energetics for luminescent (Fig.~\ref{fig5}f) and 
non-luminescent (Fig.~\ref{fig3}d) cases. Experimentally, there is a 
shift in the position of the dI/dV peak: around 1.4 V for the 
non-luminescent case and 2.0 V for the luminescent case (see 
Fig.~\ref{fig4})). A similar shift is found theoretically with Approach 
A, due to the fact that the HOMO-LUMO1 energy difference in the 
luminescent case (Fig.~\ref{fig5}e) is greater than the HOMO-LUMO energy 
difference in the non-luminescent case (Fig.~\ref{fig3}c).

The physical reason for this difference between Approach A and Approach 
B is that in Approach A the molecule-substrate coupling splits the LUMO 
degeneracy in the luminescent case but not in the non-luminescent case 
and this difference in electronic structure results in the different 
bias voltages at which the high bias peak in dI/dV occurs. By contrast, 
in Approach B the LUMO degeneracy is lifted in both the luminescent and 
non-luminescent cases, so that the electronic structure of the molecule 
and the bias voltage at which the high bias peak in dI/dV occurs is 
similar in the two cases.

\subsubsection{Dependence on the zero-bias Fermi level} 
Experimentally, different dI/dV curves are observed depending on the 
location of the molecule on the substrate (see 
Fig.~\ref{fig4})\cite{Ho03}. Even among those molecules that luminesced 
(A and B), there are differences in dI/dV. It should be noted that, in 
our article, we have chosen a zero-bias Fermi level of -10.1 eV, and 
that variations in the Fermi level relative to the molecular levels at 
zero bias are likely, depending on the location of the molecule on the 
surface, due to local work function variations (discussed in Sec. IIC). 
The dashed line in Fig.~\ref{fig5}d shows an I-V curve (using Approach 
B) for an alternate zero-bias value of $E_F$: -10.3 eV instead of -10.1 
eV. Here, the low-bias dI/dV peak is at 0.5 eV, corresponding more 
closely to Fig.~\ref{fig4}A than Fig.~\ref{fig4}B. It is possible that 
the experimental differences in low-bias dI/dV peak locations in Qiu's 
Fig.2A and Fig.2B are due to different zero-bias Fermi levels caused by 
local work function variations on the surface. Other than the change in 
the low-bias dI/dV peak location, small changes in the Fermi level yield 
qualitatively similar I-V and photon emission results. Therefore, in the 
rest of this article, we have assumed a Fermi level of -10.1 eV.

\subsection{Weak fourfold-symmetric molecule-substrate coupling}

The final case we consider is weak molecule-substrate coupling, where 
the molecule-substrate interaction is fourfold-symmetric, along with 
stronger tip-molecule coupling than in the previous cases. In this case, 
the electronic molecule-substrate coupling is of the same order of 
magnitude as the tip-molecule coupling.\cite{weakfoot} This situation 
may be achieved experimentally by increasing the thickness of the oxide 
layer between the molecule and metal substrate by a modest amount, or by 
decreasing the tip-molecule distance. In our model, we both increase the 
molecule-substrate distance and decrease the tip-molecule distance. 

\subsubsection{Approach A}
As in Sec.IIIA(1), we assume there is no splitting of the LUMO 
degeneracy. For this case, much more efficient photon emission is 
predicted to occur, with a photon yield two orders of magnitude higher 
than for Sec. IIIB. Fig.~\ref{fig8}a shows the emission spectrum at high 
bias ($V_{bias}=1.95$ V). The peak in the spectrum corresponds to the 
HOMO-LUMO transition. Fig.~\ref{fig8}c shows the I-V curve for this 
case. There is a high-bias dI/dV peak (at 1.45 V) and no low-bias peak.

Looking at Fig.~\ref{fig8}e, the molecular orbital energetics are 
similar to those for Sec.IIIA(1) (shown in Fig.~\ref{fig3}c). Since no 
orbitals enter the Fermi energy window at low bias, there is no low-bias 
dI/dV peak. In this case, the I-V curve is quite flat up to about 1.4 V. 
At 1.4 V, the LUMO fully enters the Fermi energy window, becoming 
partially occupied. The HOMO depopulates by an equal amount, and the 
orbitals electrostatically shift downwards in energy with $\mu_S$.

Since the tip has a much stronger effect on the LUMO occupation in this 
case than in Sec.IIIA,B, the degree of partial population of the LUMO, 
and partial depopulation of the HOMO, is much greater. This results in 
much greater quantum efficiency for photon emission. Unlike in Sec.IIIB, 
the initial onset voltage for photon emission due to HOMO-LUMO 
transitions matches the HOMO-LUMO emission peak energy.

\subsubsection{Approach B}
As with Approach A, with Approach B very strong photon emission is 
predicted to occur. Fig.~\ref{fig8}b shows the emission spectrum at high 
bias ($V_{bias}=1.80$ V). Unlike for Approach A, here there are {\it 
two} peaks in the spectrum, corresponding to HOMO-LUMO1 and HOMO-LUMO2 
transitions. Fig.~\ref{fig8}d shows the I-V curve for this case. There 
are two high-bias dI/dV peaks (at 1.2 V and 1.6 V) and no low-bias peak.

These results can be understood by studying the behaviour of the 
molecular orbitals with applied bias voltage (see Fig.~\ref{fig8}f). At 
low bias, the Fermi energy window approaches the LUMO1 and LUMO2. Unlike 
the other cases (Sec.IIIA,B), in this case the LUMO2 coupling to tip and 
substrate is not strongly asymmetric, and electron states from the tip 
have a significant effect on the charge of the orbital; therefore, the 
LUMO2 electrostatically shifts in energy with $\mu_T$ so that the 
zero-bias charge on the molecule is maintained. Since no orbitals enter 
the Fermi energy window at low bias, there is no low-bias dI/dV peak. At 
1.2 V, the HOMO energy reaches $\mu_S$, and the HOMO begins to 
depopulate. This causes an electrostatic shift in orbital energy 
downwards, and the LUMO2 enters the Fermi energy window, creating a 
dI/dV peak at 1.2 V. At 1.6 V, the LUMO1 enters the Fermi energy window, 
resulting in another dI/dV peak. (This increase in current is greater 
than the increase at 1.2 V, because the LUMO1 has stronger electronic 
coupling than the LUMO2 to the tip probe.) The HOMO depopulates 
significantly further, with the LUMO1 populating by an equal amount. 
(The resulting electrostatic deviation in orbital energies is too small 
to be visible in Fig.~\ref{fig8}f because the occupation of the HOMO is 
very sensitive to any deviation in energy away from $\mu_S$.)

As with Approach A, the result is higher quantum efficiency for photon 
emission. Unlike in Sec.IIIB(2), the initial onset voltage for photon 
emission due to HOMO-LUMO2 transitions corresponds to the HOMO-LUMO2 
emission peak energy. The HOMO-LUMO2 emission peak increases further 
once the onset voltage corresponding to the HOMO-LUMO1 emission peak is 
reached (due to the further depopulation of the HOMO).

A signature of this relatively efficient photon emission regime, found 
with both Approach A and Approach B, is the {\it lack} of a low-bias 
dI/dV peak. This regime has yet to be realized in STM experiments; 
however, it is predicted that greatly enhanced quantum efficiency could 
be achieved by further weakening the coupling of the molecule to the 
metallic substrate, or by bringing the STM tip closer to the molecule. 
While in our model we both increase the molecule-substrate distance and 
decrease the tip-molecule distance, it may be more experimentally 
feasible to increase the thickness of the oxide layer without bringing 
the tip closer to the molecule. This would cause a reduced current 
through the molecule. For such an experimental situation, the relevant 
luminescence observation is not the absolute photon emission intensity, 
but the quantum efficiency, or {\it photon yield} (the number of photons 
given off per electron passing through the molecule). This is predicted 
to be greatly enhanced.

\subsection{Discussion of Results}

Both Approach A and Approach B yield results consistent with experiment. 
For the case where the molecule is strongly coupled to the substrate, 
very weak photon emission, along with only a single high-bias dI/dV 
peak, is found with both approaches. Experimentally, all molecules that 
did not luminesce had a single high-bias dI/dV peak signature and no low 
bias dI/dV peak.

For the case where only a localized region of the molecule is strongly 
coupled to the substrate, both approaches yield much stronger photon 
emission than the first case. This is because, for a HOMO-LUMO 
transition, the relevant coupling asymmetry (between the tip-LUMO and 
the HOMO-substrate) is greatly reduced. Two emission peaks were found, 
the lower-energy peak being significantly lower in energy at onset than 
the energy corresponding to the onset voltage. As well, in this case 
both a low-bias and high-bias dI/dV peak are found. This is consistent 
with experiment: In the experimental case where both low-bias and 
high-bias dI/dV peaks are observed, photon emission is also observed. 
Furthermore, there is additional evidence based on modelling of the 
molecular STM images\cite{Buker05} that this experimental case 
corresponds to a localized region of strong coupling of the molecule to 
the substrate.

One qualitative feature observed experimentally and found theoretically 
with Approach A is not found with Approach B: Experimentally, there is a 
shift in the position of the high-bias dI/dV peak, depending on whether 
or not photon emission is observed. This shift is predicted with 
Approach A but not with Approach B.

There is additional experimental evidence in support of Approach A in 
the form of an observed zero-bias splitting in the LUMO degeneracy of a 
similar molecule (magnesium porphine) above the same 
Al$_2$O$_3$/NiAl(110) substrate.\cite{Wu06} It should be noted that for 
this experiment, only MgP molecules with two-lobe STM images were chosen 
for detailed study, so the substrate-dependence of the zero-bias 
splitting is unknown.

There is, however, a possible physical justification for Approach B. 
When a bias voltage is applied, the STM tip will electrostatically 
affect different molecular orbitals differently. The extent of these 
different effects is unknown. A simple electrostatic calculation, 
treating the tip/substrate as a point charge and a mirror image charge, 
suggests small differences (typically on the order of 100ths of an eV) 
in the average potential for the LUMO1 and LUMO2 orbitals. Thus, while 
the assumptions for Approach B may indeed be qualitatively correct, the 
degree to which the orbital energies of the LUMO1 and LUMO2 behave 
differently with bias is unknown and may be small.

For the case of very weak molecule-substrate coupling, much higher 
quantum efficiencies for photon emission are predicted to occur. This 
regime has not yet been realized experimentally, and would be an 
intriguing avenue for further research.

\section{Conclusions}
\label{Conclusions}

The local-electrode framework presented in this article coherently 
explains a multitude of experimental observations\cite{Ho03} not 
previously theoretically studied, for the 
STM/Zn-etioporphyrin/Al$_2$O$_3$/NiAl(110) system. The following is a 
summary of these observations, with explanations based on our model 
results:

\renewcommand{\labelenumi}{(\roman{enumi})}
\begin{enumerate}

\item{The observed molecular-based photon emission is due to transitions 
between the molecular LUMO, whose degeneracy has been split by 
molecule-substrate and/or molecule-STM tip interactions, and the 
molecular HOMO.}

\item{For some cases, low-bias dI/dV peaks are observed experimentally 
(see Fig.~\ref{fig4}A,B). Our model explains these as being due to a 
splitting of the LUMO degeneracy, with the lower-energy LUMO entering 
the Fermi energy window at low bias (see Sec. IIIB).}

\item{For some cases, no low-bias dI/dV peak is experimentally observed 
(see Fig.~\ref{fig4}C-F). We find that this occurs because the molecule 
is too strongly coupled to the substrate, with the LUMO either not 
entering the Fermi energy window at low bias (Approach A, see 
Sec.IIIA(1)), or entering the window but contributing negligibly to the 
current due to very weak coupling of the molecule to the tip compared to 
the substrate (Approach B, see Sec. IIIA(2)).}

\item{For cases with no low-bias peak, no photon emission is 
experimentally observed. This is due to strongly asymmetric tip/molecule 
and molecule/substrate couplings. In these cases, when a bias is 
applied, the HOMO stays almost fully occupied and the LUMO almost 
completely unoccupied (see Sec. IIIA(1,2)).}

\item{There is an experimentally observed difference in the position of 
the high-bias dI/dV peak, between cases where photon emission is and is 
not observed (see Fig.~\ref{fig4}). This is explained with Approach A by 
a breaking of the LUMO degeneracy only in the luminescent case (see Sec. 
IIIA(1) and Sec. IIIB(1)).}

\item{The experimental peak photon energy is about 0.5 eV below 
$eV_{bias}$ at emission onset (see Fig.~\ref{fig7}). This is due to 
splitting of the LUMO degeneracy, with the lower-energy LUMO being well 
inside the Fermi energy window as the energy of the HOMO approaches the 
window. See Sec. IIIB(1,2).}

\item{The high-energy photon emission spectra tails are steeper than the 
low-energy tails (see Fig~\ref{fig7}). This is due to the substrate 
Fermi energy providing a sharp energy cutoff below which there are no 
available states to receive a transition (see Sec. IIIB(1,2)).}

\item{There are significant differences in experimentally observed 
positions of dI/dV peaks (see Fig~\ref{fig4}) depending on the position 
of the molecule on the substrate. These differences are consistent with 
differing local zero-bias Fermi levels, due to local variations in the 
work function of the oxide-coated metal substrate (see Sec. IIIB(3)).}

\end{enumerate} 

Our model predicts an additional photon emission peak to be found, 
having a peak energy close to the bias voltage at emission onset, for 
the case of molecular-based photon emission presented in Sec. IIIB. 
Experiments testing this prediction would be of interest.

We also predict that greatly enhanced quantum efficiency of photon 
emission could be achieved by further weakening the coupling of the 
molecule to the metallic substrate, or if possible by bringing the STM 
tip closer to the molecule (see Sec. IIIC). The spectrum for this 
greatly enhanced quantum efficiency could yield further clues to the 
relative merits of the two approaches studied in this article (see 
Sec.IID). For Approach A, one emission peak is predicted, and for 
Approach B, two peaks are predicted.

Studying the STM/Zn-etioporphyrin/Al$_2$O$_3$/NiAl(110) system using the 
local-electrode theoretical framework presented in this paper has 
yielded a coherent explanation of a large body of experimental results 
for this system. Using this framework, we are able to gain a much 
greater understanding of single molecule electroluminescence. This is an 
important step towards the development of the emerging field of single 
molecule optoelectronics. We hope that this work inspires further 
experimental and theoretical research in this promising new field.

While the present theory relies heavily on phenomenology, it has allowed 
us to construct energy level diagrams of the evolution of the molecular 
HOMO and LUMO orbitals and of the electrochemical potentials of the 
electrodes as a function of applied bias that are physically reasonable 
and are consistent with {\em both} the experimentally observed 
current-voltage characteristics {\em and} the experimental 
electroluminescence data. Thus the present work can also be viewed as a 
quantitative interpretation of the experimental data that is unique in 
that it satisfies more demanding experimental constraints than previous 
attempts to model experimental molecular electronic data that have 
focussed on experimental current-voltage characteristics alone. 
Therefore, as well as contributing to a better understanding of 
single-molecule optoelectronics the present work provides much needed 
benchmarks for the development of accurate first principles theories of 
the evolution of the electronic structure of molecular nanowires under 
bias that do not yet exist at this time.

\newpage

\begin{acknowledgments}
We are grateful to Brad Johnson and Ross Hill for helpful discussions. 
This research was supported by NSERC. G.~K.~ is grateful for support in 
the form of a Fellowship from the Canadian Institute for Advanced 
Research. Some numerical computations presented in this work were 
performed on WestGrid computing resources, which are funded in part by 
the Canada Foundation for Innovation, Alberta Innovation and Science, BC 
Advanced Education, and the participating research institutions. 
WestGrid equipment is provided by IBM, Hewlett Packard, and SGI. 
\end{acknowledgments}

\newpage

\begin{figure}[!t]
\caption{(Color Online) A schematic energy level diagram of a transition 
from an occupied electron state (incoming from the tip probe on the 
left, shown in red) to an unoccupied electron state (incoming from a 
substrate contact on the right, shown in blue). A photon is created with 
energy $\hbar \omega$, equal to the difference in energy between the two 
electron states. The dashed lines represent the molecular portions of 
the states.}
\label{fig1}
\end{figure}

\begin{figure}[!t]
\caption{(Color Online) The Zn(II)-etioporphyrin I molecule, showing 
substrate contacts S$_1$, S$_2$, S$_3$, and S$_4$ (open blue circles, 
into the page) and the tip probe (blue dot, out of page). Carbon atoms 
are red, nitrogen atoms are green, the zinc atom is yellow, and hydrogen 
atoms are white.}
\label{fig2}
\end{figure} 

\begin{figure}[!t]
\caption{(Color Online) Strong fourfold-symmetric coupling between 
molecule and substrate: Electric current, and molecular orbital energies 
as a function of bias voltage. (a) Approach A, I vs. $V_{bias}$. Red 
lines represent dI/dV. (b) Approach B, I vs. $V_{bias}$. (c) Approach A, 
molecular orbital energies (dashed lines represent tip and substrate 
electrochemical potentials). (d) Approach B, molecular orbital 
energies.}
\label{fig3}
\end{figure} 

\begin{figure}[!t] 
\caption{(Color Online) From Qiu et al.\cite{Ho03}. Reprinted with 
permission from AAAS. Experimental dI/dV curves for 
Zn-etioporphyrin/Al$_2$O$_3$/NiAl(110) obtained with the STM, for 
molecules at different locations on the substrate. (A-F) dI/dV curves 
representative of the various molecular images observed. The curve seen 
in B was most commonly observed (30\% of the time). Molecular 
electroluminescence was observed for cases A and B but not for C-F.
}
\label{fig4}
\end{figure}

\begin{figure}[!t]
\caption{(Color Online) Localized strong coupling: photon emission, 
electric current, and molecular orbital energies as a function of bias 
voltage. (a) Approach A, photon emission vs. $V_{bias}$. (b) Approach B, 
photon emission vs. $V_{bias}$. (c) Approach A, I vs. $V_{bias}$. Red 
lines represent dI/dV. (d) Approach B, I vs. $V_{bias}$. Dashed line 
represents $E_F=-10.3$ eV. (e) Approach A, molecular orbital energies. 
(f) Approach B, molecular orbital energies.}
\label{fig5}
\end{figure} 

\begin{figure}[!t]
\caption{Localized strong coupling: Onset of photon emission, at the 
HOMO-LUMO2 emission peak. (a) Approach A, emission rate vs. photon 
energy, for three different values of $V_{bias}$ around the onset 
voltage. (b) Approach B, emission rate vs. photon energy.}
\label{fig6}
\end{figure} 

\begin{figure}[!t]
\caption{(Color Online) From Qiu et al.\cite{Ho03}. Reprinted with 
permission from AAAS. Experimental photon emission spectra for molecules 
corresponding to Fig.\ref{fig4}B, for various bias voltages around the 
onset voltage.}
\label{fig7}
\end{figure}

\begin{figure}[!t]
\caption{Weak fourfold-symmetric coupling: photon emission, electric 
current, and molecular orbital energies as a function of bias voltage. 
(a) Approach A, photon emission vs. $V_{bias}$. (b) Approach B, photon
emission vs. $V_{bias}$. (c) Approach A, I vs. $V_{bias}$. (d) Approach
B, I vs. $V_{bias}$. (e) Approach A, molecular orbital energies. (f) 
Approach B, molecular orbital energies.}
\label{fig8}
\end{figure}


\begin{references}
\bibitem{plasmonfoot} Some of the data in Ref. \onlinecite{Ho03} 
involves Zn-etioporphyrin on bare NiAl(110), with no insulating 'spacer' 
layer. For this data, plasmon-based photon emission is significant and 
molecular-based photon emission is quenched. The model in the present 
article has been created to study systems involving an insulating 
'spacer' layer. Therefore, our discussion of the data in Ref. 
\onlinecite{Ho03} is limited to the cases with a 'spacer' layer, where 
plasmon-based emission was shown to be negligible 
experimentally,\cite{Ho03} and plasmon effects are not considered in 
this article.
\bibitem{vcancel} For 
numerical convenience we have chosen 
the site energies $\epsilon$ and hopping
amplitudes
$\beta$ in Eq. (\ref{Helectrodes}) to be the same for all of the
ideal one-dimensional tight-binding leads that represent the
electrodes, and to be independent of the applied bias voltage.
Hence in Eq.(\ref{Landauercurrent})
$T=\sum_{i=1}^m\frac{v^\prime}{v}|t_i|^2=\sum_{i=1}^m|\psi_{1,i}|^2$ 
since the velocities of the incoming electron ($v$) and 
the (elastically scattered) transmitted electron ($v^\prime$) are equal.
However, the transmission probability $T$ 
depends on the applied bias voltage $V_{bias}$ due 
to the dependence of the molecular orbital 
energies on the bias that is calculated as described in Sec. IID.
\bibitem{strongfoot} In order to simulate strong 
molecule-substrate
coupling on a uniform oxide surface, each substrate contact in 
this case
is positioned 4.0 \AA below the plane of the molecule (2.5\AA 
 from the
nearest atom of the molecule). The tip probe is positioned 4.1 
\AA above
the molecule. See Fig.\ref{fig2}.
\bibitem{RatioNote} Changing the ratio 
$\alpha:\gamma_1:\gamma_2$   
affects the energies of the relevant molecular orbitals when a 
bias
voltage is applied, but does not qualitatively change the model 
results
as long as the condition $\gamma_2 < \alpha < \gamma_1$ is met. 
For a
ratio $\alpha:\gamma_1:\gamma_2$ set to 9:10:8 instead of 3:4:2
(corresponding to smaller differences in the electrostatic 
effects on
each orbital), all of the qualitative features described in 
Sec.III 
remain the same.
\bibitem{strongmodfoot} Here, the substrate contact $S_4$ is 
positioned 
4.0 \AA below the plane of the molecule, with the other three 
substrate
contacts positioned 4.5 \AA below the plane. The tip probe is 
positioned
4.5 \AA above the molecule.
\bibitem{splitfoot} Since the LUMO2 is strongly coupled to the
substrate (similar to case A: Strong fourfold-symmetric
molecule-substrate coupling), we assume the splitting of 
degeneracy
primarily changes the LUMO1 energy:
$\epsilon_{LUMO1}=\epsilon_{LUMO}+0.5$ eV,
$\epsilon_{LUMO2}=\epsilon_{LUMO}$.
\bibitem{weakfoot} Here, the substrate contacts are positioned 
4.5 \AA
below the plane of the molecule (3.0 \AA from the nearest atom 
of the
molecule) and the tip probe is positioned 4.0 \AA above the 
molecule.
The lateral electrode positions are the same as for the other 
cases (see
Fig.\ref{fig2}).
\end{references}
\end{document}